%% file: conference_101719.tex
\documentclass[sigconf]{acmart}

\AtBeginDocument{%
  \providecommand\BibTeX{{%
    \normalfont B\kern-0.5em{\scshape i\kern-0.25em b}\kern-0.8em\TeX}}}

\copyrightyear{2022} 
\acmYear{2022} 
\setcopyright{acmcopyright}\acmConference[ICSE '22]{44th International Conference on Software Engineering}{May 21--29, 2022}{Pittsburgh, PA, USA}
\acmBooktitle{44th International Conference on Software Engineering (ICSE '22), May 21--29, 2022, Pittsburgh, PA, USA}
\acmPrice{15.00}
\acmDOI{10.1145/3510003.3510160}
\acmISBN{978-1-4503-9221-1/22/05}

\usepackage{CJKutf8}
\usepackage{svg}
\usepackage{tcolorbox}
\usepackage{fancybox}
\usepackage{bm}
\usepackage{multirow}

\usepackage{natbib}
\setcitestyle{square, comma, numbers,sort&compress, super}
\usepackage{amsmath,amsfonts}
\usepackage{algorithmic}
\usepackage{graphicx}
\usepackage{textcomp}
\usepackage{xcolor}
\usepackage{xspace}
\usepackage{flushend}
\usepackage{soul}
\usepackage{hyperref}
\usepackage[capitalise]{cleveref}
\usepackage{enumitem}
\usepackage{wrapfig}
\usepackage{caption}
\usepackage[export]{adjustbox}
\usepackage{subcaption}

\newcommand\numberthis{\addtocounter{equation}{1}\tag{\theequation}}
\def\BibTeX{{\rm B\kern-.05em{\sc i\kern-.025em b}\kern-.08em
    T\kern-.1667em\lower.7ex\hbox{E}\kern-.125emX}}

\soulregister\cite7 
\newcommand{\tool}{\emph{NLQF}\xspace}
\newcommand{\dataset}{\emph{COFIC}\xspace}
\newcommand{\corpus}{bootstrap query corpus\xspace}

\begin{document}

\title{On the Importance of Building High-quality Training Datasets for Neural Code Search}

\author{Zhensu Sun}
\email{zhensuuu@gmail.com}
\affiliation{%
  \institution{Monash University}
  \city{Melbourne}
  \state{Victoria}
  \country{Australia}
}

\author{Li Li}
\email{1853549@tongji.edu.cn}
\affiliation{%
  \institution{Tongji University}
  \city{Shanghai}
  \country{China}
}

\author{Yan Liu}
\email{yanliu.sse@tongji.edu.cn}
\affiliation{%
  \institution{Tongji University}
  \city{Shanghai}
  \country{China}
}

\author{Xiaoning Du}
\authornote{Xiaoning Du and Li Li are co-corresponding authors.}
\email{xiaoning.du@monash.edu}
\affiliation{%
  \institution{Monash University}
  \city{Melbourne}
  \state{Victoria}
  \country{Australia}
}

\author{Li Li}
\authornotemark[1]
\email{li.li@monash.edu}
\affiliation{%
  \institution{Monash University}
  \city{Melbourne}
  \state{Victoria}
  \country{Australia}
}

\begin{abstract}
The performance of neural code search is significantly influenced by the quality of the training data from which the neural models are derived.
A large corpus of high-quality query and code pairs is demanded to establish a precise mapping from the natural language to the programming language.
Due to the limited availability, most widely-used code search datasets are established with compromise, such as using code comments as a replacement of queries.
Our empirical study on a famous code search dataset reveals that over one-third of its queries contain noises that make them deviate from natural user queries.
Models trained through noisy data are faced with severe performance degradation when applied in real-world scenarios.
To improve the dataset quality and make the queries of its samples semantically identical to real user queries is critical for the practical usability of neural code search.
In this paper, we propose a data cleaning framework consisting of two subsequent filters: a rule-based syntactic filter and a model-based semantic filter. 
This is the first framework that applies semantic query cleaning to code search datasets.
Experimentally, we evaluated the effectiveness of our framework on two widely-used code search models and three manually-annotated code retrieval benchmarks.
Training the popular DeepCS model with the filtered dataset from our framework improves its performance by 19.2\% MRR and 21.3\% Answer@1, on average with the three validation benchmarks. 
\end{abstract}



\maketitle

\input{intro}

\input{preliminary}
\input{approach}
\input{experiments}
\input{application}
\input{threatstovalidity}

\input{relatedwork}

\section{Conclusion}
\label{sec:conclusion}
We propose the first data cleaning framework for code search tasks, which improves the quality and naturalness of the queries. 
The framework leverages two subsequent filters, the rule-based syntactic filter, and the model-based semantic filter.
The rule-based filter uses configurable heuristics rules to filter out comments with syntactic anomalies. 
The model-based filter aims to refine the dataset semantically.
It trains a VAE model over a pre-collected \corpus, and exploits it to select comments with smaller reconstruction losses. Experiments show that our filtering framework can significantly save computing resources and improve the model accuracy. 
Finally, we release our framework as a Python library \tool and make public a high-quality cleaned code search dataset \dataset, to facilitate relevant research in academia and industry.


\flushend
\bibliographystyle{ACM-Reference-Format}
\bibliography{base}

\end{document}

%% file: intro.tex
\section{Introduction}
\label{sec:intro}

A semantic code search engine is a vital software development assistant, which significantly improves the development efficiency and quality. 
With a description of the intended code functionality in natural language, a search engine can retrieve a list of semantically best-matched code snippets from its codebase. 
Recently, deep learning (DL) has been widely applied in this area in view of its advantages in semantic modeling and understanding of languages. In the task of code search, DL models learn and represent the semantic mappings between the natural language and the programming language from query-code pairs.

\begin{figure}[t]
\centerline{\includegraphics[width=0.96\columnwidth]{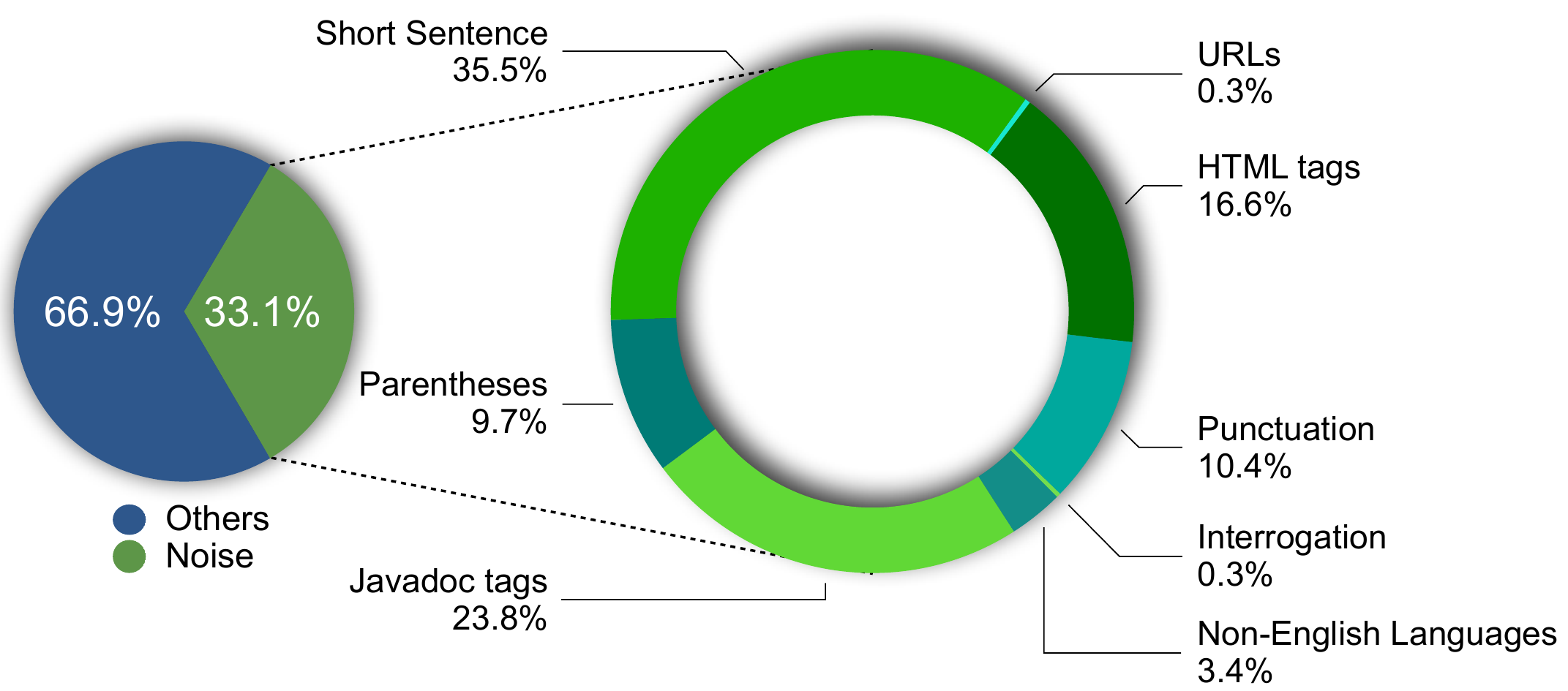}}
\caption{Statistics of 394,471 code comments used in CodeSearchNet (Java). The feature definitions are presented in \cref{sec:rule}.}
\label{fig:statistic}
\end{figure}

Like many other DL tasks, code search models are data-hungry and require large-scale and high-quality training datasets.
Nevertheless, collecting a large set of query-code pairs is challenging, where the queries are supposed to be natural expressions from developers and the code to be a valid semantic match.
Instead, considering the scale and availability, code comments are popularly used as an alternative to the queries, many of which describe the core functionalities and with the corresponding code implementation rightly available.
To better understand the quality of datasets hence constructed, we investigated a Github dataset, CodeSearchNet (Java)~\cite{Husain2019CodeSearchNetCE},  which is popularly used in current code search research.
Surprisingly, we found a considerable amount of noise and unnaturalness in the queries of its data samples, which can hinder the training of high-quality models for practical usage. 
As shown in~\cref{fig:statistic}, one-third of its queries contain text features (see Table~\ref{tab:definations} for examples of different features) that hardly exist in actual user queries.
The features are summarized based on our observations of the dataset, and may not be sufficient. 
Comments may also be used for other purposes, such as copyright and to-do, instead of describing the core functionalities, thus shall not be seen as queries.
The proportion of noise data can be higher than one-third.

Code search models trained with noisy queries will face severe performance degradation when dealing with actual user queries.
The gap between the collected comment-code pairs and the natural user queries violates the basic assumption of learning algorithms that the training data and the evaluation data share a similar distribution.
It is also noteworthy that evaluating the model with a noisy comment-code benchmark can hardly reflect how useful the model would be in practice, and, even worse, may bring non-negligible bias to the model design, evaluation and application.
Many other researchers~\cite{Cambronero2019WhenDL,Ling2020AdaptiveDC,sun2020pscs} also point out the misalignment between code comments and natural user queries, and report it as a threat to the validity of their approaches.
As mentioned in~\cite{Liu2020OpportunitiesAC, zhao2021impact}, improving the quality of the training data is still a research opportunity for machine learning, including DL-based code search models. 
Considering that there are still plenty of comments close to actual user queries and naturally paired with high-quality code snippets, a promising solution is to filter out the noisy ones.
Manual filtering can produce the most accurate results but is hardly practical for large-scale datasets.
Automated data cleaning methods are of demanding needs.

Queries for code retrieval possess specific \emph{syntactic} and \emph{semantic} characteristics, which can be utilized as key features to distinguish genuine user queries from noise. 
Typical syntactic features include text attributes such as keywords, sentence length, and language type. Semantic features are related to the intention underneath the text expression, which usually describes the computational functionality of code snippets and might be influenced by the design convention of common program APIs. 
Compared with syntactic features, semantic features are more abstract, implicit, and hard to be matched by simple rules.
Recently, some initial efforts have emerged on query quality enhancement, but primarily focusing on the regularization of syntactic features.
Simple filtering heuristics are proposed, based on the appearance of verb and noun phrases~\cite{Ling2020AdaptiveDC}, keywords uncommonly used in queries~\cite{Cambronero2019WhenDL}, and constraints on the query length~\cite{Husain2019CodeSearchNetCE,Ling2020AdaptiveDC}.

However, the improvement in data quality is limited.
As declared in~\cite{Husain2019CodeSearchNetCE}, the collected dataset is still noisy despite their data cleaning efforts. 
The proposed rules are not sufficient to cover the various syntactic violations, let alone the semantic misalignment.
For example, warning messages such as ``Use of this property requires Java 6'' widely exist in the code comments, but few code queries would request this way.
Hence, a remaining challenge is recognizing the code comments that are syntactically valid but encode semantics rarely seen in natural user queries.

To tackle this challenge, we propose an automated and effective data cleaning framework that distills high-quality queries from generally collected code comments on a large scale.
The framework is orthogonal to the design of code search algorithms and could be integrated with any of them to improve the quality of the training dataset. 
Basically, it encompasses two subsequent filters: \emph{a rule-based syntactic filter} and \emph{a model-based semantic filter}. 
The rule-based filter includes a set of systematically designed heuristic rules and weeds out data with anomalous syntactic features, e.g., HTML tags and Javadoc tags. 
It is developed to cover a diverse range of syntactic violations, and each member inside is validated to reduce the noises effectively.
It is also extensible to fulfill the specific requirements for the dataset based on the applications. 
The model-based semantic filter further refines the dataset produced by the rule-based filter and retains the comments that are semantically close to the natural queries.
The filter relies on a \corpus, a set of high-quality queries, which represents how semantically the queries should look.
It learns the semantic features of the corpus, such as the expression style and topic, with a DL model, and leverages it to identify samples with similar semantics. 
The \corpus could be constructed with any trusted sources of natural user queries, and we formulate it with question titles from StackOverflow in this work.
These titles are an ideal approximation of natural queries and could be re-used by related studies.
Then, a Variational Auto-Encoder~\cite{Kingma2014AutoEncodingVB} is trained with the \corpus, which maps the inputs into a latent space and attempts to reconstruct the original inputs solely based on the latent features.
The reconstruction loss reflects ``how far away'' an input is from the training data distribution, i.e., the distribution of queries in the \corpus.
The lower the reconstruction loss, the more qualified an input is as a natural query.
We compute the reconstruction loss for each code comment in the raw dataset and cluster them into two groups.
The group of qualified queries is retained for training, and the group of noises is discarded.

\begin{table}[t]
\caption{Examples of syntactic rules.}
\begin{center}
\input{tables/definitions}
\end{center}
\label{tab:definations}
\end{table}

To evaluate the effectiveness of our data cleaning framework, we compare the performance of code search models trained with datasets before and after the filtering. 
One training dataset, two neural models, and three manually annotated validation datasets are used in the experiments, and our framework brings a significant performance improvement under all settings.
In particular, the performance of the popular DeepCS~\cite{Gu2018DeepCS} model is improved by 19.2\% MRR and 21.3\% Answer@1, on average with the three validation datasets. 
More importantly, with less training data used after the filtering, we also save the training time and computation resources.
Further, we carry out a comprehensive ablation study to validate the usefulness of each filter component and each rule and manually inspect the quality of the rejected and retained data.
Finally, we release the implementation of our framework, \tool, and a cleaned code search dataset, \dataset, to facilitate future research. 
The source code and datasets are available at \url{https://github.com/v587su/NLQF}.

To the best of our knowledge, this is the first systematic data cleaning framework for comment-based code search datasets. Our main contributions include:

\begin{itemize}[leftmargin=*]
\item A two-step data cleaning framework for code search datasets, which bridges the gap between code comments and natural user queries, both syntactically and semantically.

\item Implementation of the framework as a Python library for the code search task in academia and industry.

\item A comprehensive evaluation of our framework's effectiveness, which demonstrates significant model performance improvement on three manually-annotated validation benchmarks.

\item The first systematically distilled Github dataset for neural code search, containing over one million comment-code pairs.
\end{itemize}


%% file: tables/definitions.tex
\begin{tabular}{|l|l|l|}
\hline
\textbf{Syntax Feature} & \textbf{Rule Action} & \textbf{Example} \\ \hline
HTML tags & Partly Remove & \textless{}p\textgreater{}parse line\textless{}/p\textgreater{} \\ \hline
Parentheses & Partly Remove & (TODO) Send requests \\ \hline
Javadoc tags & Fully Remove & Returns a \{@link Support\} \\ \hline
URLs & Fully Remove & See https://github.com/ \\ \hline
\begin{tabular}[c]{@{}l@{}}Non-English\\ Languages\end{tabular} & Fully Remove & \begin{CJK}{UTF8}{gbsn}创建临时文件\end{CJK} \\ \hline
Punctuation & Fully Remove & ============== \\ \hline
Interrogation & Fully Remove & Is this a name declaration? \\ \hline
Short Sentence & Fully Remove & DEPRECATED \\ \hline
\end{tabular}

%% file: preliminary.tex
\section{Preliminaries}
\label{sec:preliminaries}
We prepare readers with the primary sources for collecting query-code pairs and the Variational Auto-Encoder, a major building block of our framework.

\subsection{Data Source}
\label{sec:data source}
A query for neural code search describes, in natural language, the functionality of the code snippets desired by users, e.g., ``convert string to JSON object''. 
The ideal data source for genuine code queries is the production data from existing neural code search engines. 
However, these queries are not publicly accessible due to privacy and business sensitivity. 
In academia, researchers use texts with similar intentions (e.g., code comments) as a replacement. 
The primary alternative data sources for semantic code search research include GitHub and StackOverflow.

\subsubsection{Github} 
Github\cite{github} is an open-source community, hosting more than 100 million repositories. It is the most popular platform for developers to share and manage their projects.
The large-scale well-maintained repositories on Github are a treasury for code reuse during development, thus naturally becoming the main retrieval source for code search tasks.
Moreover, mature projects are usually accompanied by canonical development documents.
According to Javadoc\cite{javadoc}, a code comment style guide, the first sentence of doc comments should be a summary sentence. Therefore, it is convenient to construct a code search dataset by collecting the code snippets paired with the first sentence of comments, forming the comment-code pairs. Javadoc-generated comments have hence been widely used in practice for various software engineering purposes~\cite{li2020cda, liu2021identifying, li2018characterising, li2016accessing} due to their large scale, ease of obtaining, and being close to 
actual use scenarios.

However, developers write comments for their software projects without considering the retrieval purposes. Not all the comments properly map to queries. 
As mentioned in~\cref{sec:intro}, the CodeSearchNet collected from Github contains plenty of anomalies that rarely exist in natural user queries.
It is not appropriate to include these comments in the dataset, and we call for more attention to be drawn to this problem.

\subsubsection{StackOverflow} StackOverflow\cite{stackoverflow} serves as a Q$\&$A community specialized for software developers. It is a rich resource of software-related questions and answers. 
When asking about codes or APIs for implementing a specific functionality, users would propose a question title to express their intention. 
These are natural user queries with valid syntax and semantics. 
Hence, researchers\cite{Yin2018LearningTM,Yao2018StaQCAS} also collect the titles of StackOverflow questions paired with proper answers containing sample code snippets, which also form the query-code pairs.
Others also evaluate their code search models with queries manually selected from StackOverflow~\cite{Cambronero2019WhenDL,Gu2018DeepCS,Liu2020SimplifyingDM}, and additional public evaluation benchmarks could be found in~\cite{Li2019NeuralCS,Yan2020AreTC}.

Compared with the Github data source, queries from StackOverflow have a significant advantage of being closer to natural user queries, but the quality of code samples is hard to guarantee.
Hence, the dataset collected from StackOverflow is still not as desired.
Nevertheless, the query corpus is valuable.
It is worth investigating whether and how it could be leveraged to improve the other query-code datasets.




\subsection{Variational Auto-Encoder}
\label{sec:vae}
Variational Auto-Encoder (VAE)~\cite{An2015VariationalAB} is a neural model that learns the distribution of a set of data. A VAE model consists of an encoder and a decoder. 
The encoder learns to map an input data $x$ into a prior distribution $p_\theta(z)$, from which a latent variable $z$ is sampled, and the decoder maps $z$ back to $\hat{x}$, a reconstruction of $x$. It is expensive to calculate $p_\theta(z)$ directly, so VAE introduces an approximate posterior $q_\phi(z|x)$. 
$\theta$ and $\phi$ are parameters of the prior and the approximate posterior.

The loss function, Evidence Lower Bound (ELBO), which seeks to maximize the likelihood of reconstructing the original data and minimize the Kullback-Leibler (KL) divergence between the actual and estimated posterior distributions, is represented as:
\begin{equation}
    \mathcal{L} = \mathbb{E}_{q_\phi(z|x)}[-logp_\theta(x|z)]
+ KL(q_\phi(z|x)||p_\theta(z)), \label{eq:elbo}
\end{equation}
where $KL$ represents the KL divergence.
Theoretically, the distributions of $q_\phi(z|x)$ and $p_\theta(z)$ can be arbitrary. In practice, the Gaussian distribution is mostly adopted.



%% file: approach.tex
\section{The Data Cleaning Framework}
\label{sec:approach}
\begin{figure}[t]
\centerline{\includegraphics[width=7cm]{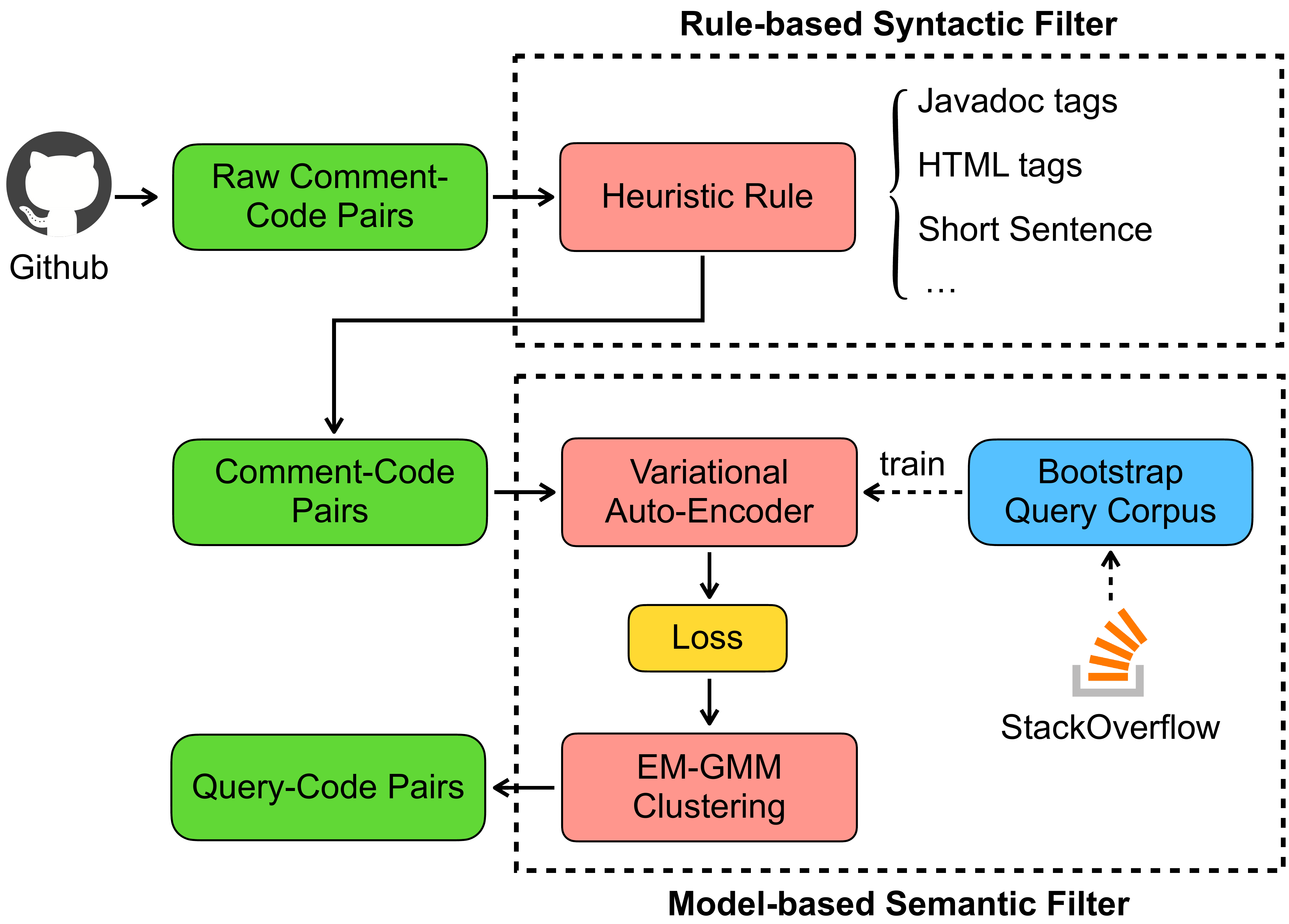}}
\caption{An overview of our data cleaning framework.}
\label{fig:process}
\end{figure}

This section introduces our automated and effective data cleaning framework for code search datasets, mainly to filter out query-code pairs with inappropriate queries.
The framework consists of two subsequent filters, the \emph{rule-based syntactic filter} and \emph{model-based semantic filter}. 
An overview of the framework, when applied to clean the comment-code pairs collected from Github, is shown in~\cref{fig:process}. 
The raw comment-code pairs are firstly cleaned by the rule-based filter, where a ruleset is applied to detect the existence of invalid query syntax. 
Next, for the model-based filter, leveraging a small \corpus as the semantics reference, a VAE model is trained to model its characteristics and further used to reject comments violating the natural query semantics. 
Here we collect the \corpus (no need of the paired code snippets) from StackOverflow, and more details can be found in~\cref{sec:dataset}.
Such, we take advantage of both the Github and StackOverflow data sources and produce a large set of high-quality query-code pairs.
In the following, we elaborate on the design of the two filters.
 
\subsection{The Rule-based Syntactic Filter}
\label{sec:rule}

Code comments contain richer information than just descriptions on code functionalities and manifest various syntactic features rarely existing in actual code search quires.
For example, URLs are used for external references, and HTML tags are used in comments for documentation autogeneration.
To reduce such deviations from natural queries, we sampled 1\% code comments from CodeSearchNet, manually inspected and summarized noises in these 3,949 instances.
We establish a black list of invalid syntax features to reject unqualified code comments.
If a comment matches any of these features, we remove the invalid parts if they are detachable; otherwise, we abandon this comment-code pair.

Based on a comparative observation of code comments and user queries, we develop a set of rules to precisely identify synthetically inappropriate queries and leave the fine-grained semantic check to the model-based filter. 
To facilitate the management, we define three criteria that the ruleset must comply with: 1) any rule should define a unique and specific construction pattern, 2) the rules should be conservative and limit the preclusion of valid queries within an acceptable range, and 3) any rule is not a subrule of other rules in the set.
As a plug-in framework, the ruleset is extensible, and any rules that meet these criteria can be appended to the set. 

We introduce the syntax features covered by our ruleset in the following, and their examples can be found in~\cref{tab:definations}.
We empirically decide whether to keep the content enclosed by a feature structure or not and validate the decisions with experiments (see our website~\cite{hqtd} for more details) and manual inspection (see~\Cref{sec:re2}).
From the results, our decisions help improve the naturalness and bring greater improvement to the model performance.

\noindent\textbf{HTML tags} HTML tags are used for documentation autogeneration in comments and should not appear in user queries. However, the content wrapped by the tags can still be informative. 
Therefore, we remove the HTML tags from the comments but keep the wrapped content.

\noindent\textbf{Parentheses} Parentheses in comments are for adding supplementary information and do not appear in user queries. Due to such purpose, the removal of the content inside the parentheses does not have much influence on the completeness of the comments. We only retain the content outside of the parentheses.

\noindent\textbf{Javadoc tags} Javadoc tags starting with an ``@'' sign are special tags indicating a Javadoc comment. Such comments are only consumed by the Javadoc project for autogenerating well-formatted documentation. Considering that the special syntax of the tags may mislead code search models on natural language understanding, we reject all comments containing Javadoc tags.
   
\noindent\textbf{URLs} URLs in comments provide external references to relevant code snippets, but natural language queries do not contain any URLs. We reject all comments containing URLs.

\noindent\textbf{Non-English Languages} Non-English expressions exist as developers from different countries may write comments in their first languages. However, current code search models are not designed to handle multi-languages. We reject all non-English comments.

\noindent\textbf{Punctuation} Sometimes, punctuation symbols are used for section partitioning in the comments.
For example, developers use a row of equal signs ($=$) or asterisks signs ($*$) (see examples in~\cref{tab:definations}) to indicate a new section.
For effectiveness, we reject comments containing no English letters in our implementation.
    
\noindent\textbf{Interrogation} Based on our observation, some of the comments in the dataset are interrogative. Developers seem to use comments to communicate with their collaborators during the code review process.
There may be some sparse information about the code functionality, but the quality is hard to control.
We reject comments ending with a question mark.
    
\noindent\textbf{Short Sentence} The sentence length is a commonly used criterion for comment filtering. 
Extremely short comments are not informative enough for code search models to establish their mapping to the corresponding code snippets.
We reject comments containing no more than two words.

\subsection{The Model-based Semantic Filter}
\label{subsection:model-filter}

This section introduces the model-based semantic filter, which takes the initially cleaned comment-code pairs from the rule-based filter as input and further selects the pairs with comments semantically close to the queries in a pre-collected \corpus.
We present the detailed design of the VAE model and discuss how it is used for filtering.

\subsubsection{The VAE Model} 
\begin{figure}[t]
\centerline{\includegraphics[width=9cm]{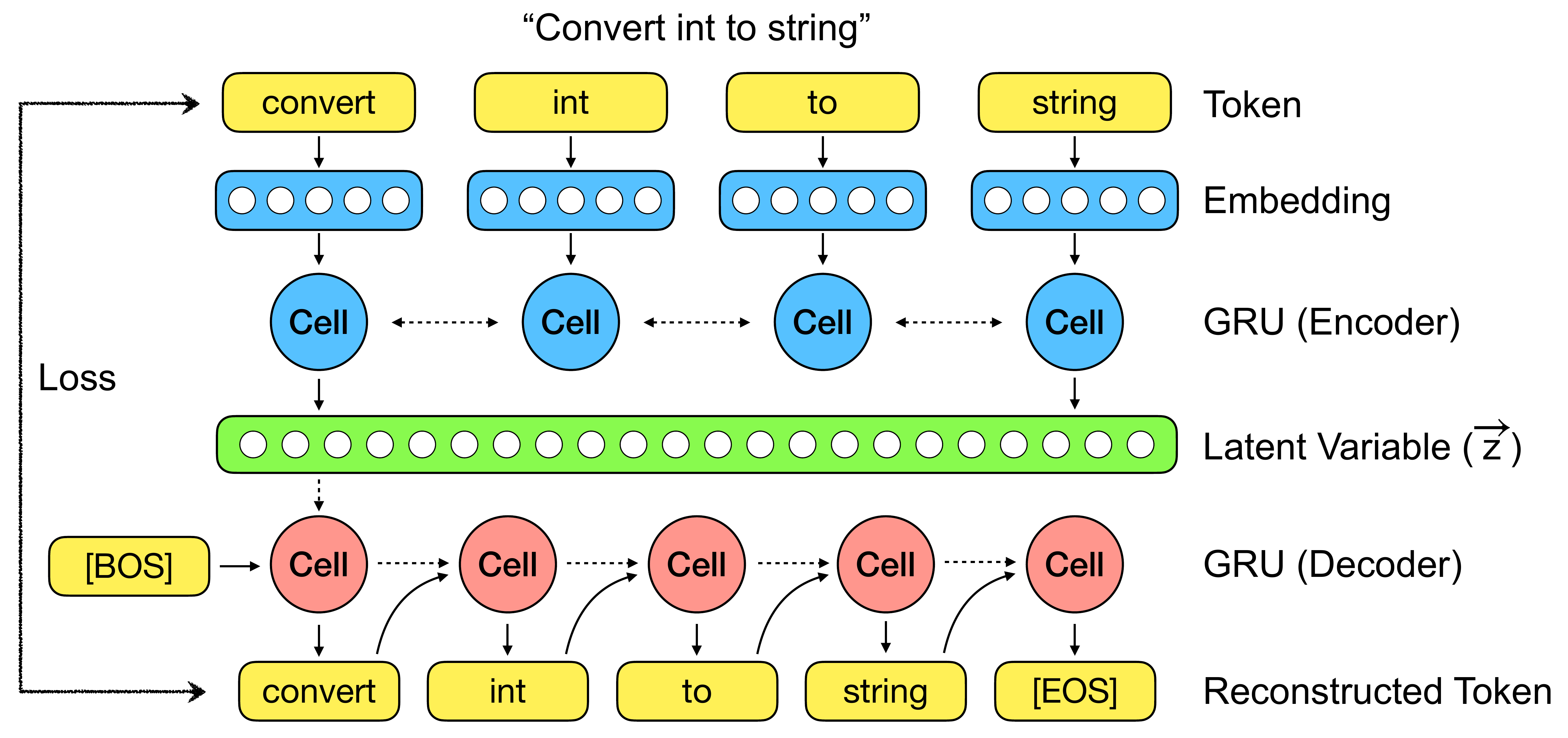}}
\caption{The structure of the Variational Auto-Encoder in the model-based filter. The dashed lines denote the propagation of hidden states in neural cells.}
\label{fig:overview}
\end{figure}

The two main components of a VAE model are the encoder and decoder, which are generally composed of deep neural networks. 
Here, we use Gated Recurrent Unit (GRU)~\cite{Cho2014LearningPR} for both the encoder and decoder in our VAE model.
GRU is a variant of Recurrent Neural Network (RNN), which enables the model to capture information from sequential texts. \cref{fig:overview} illustrates an overview of the design of the model structure. Details about each layer are as follows.

\textbf{Embedding} Given a query $w_0w_1\dots w_n$ of length $n$, the $i$-th token is $w_i$.
The embedding layer is responsible for mapping each token into an embedding vector.
It consists of an embedding matrix $\textbf{E} \in \mathbb{R}^{o_w \times d}$, where $o_w$ is the vocabulary size of the query language and $d$ is the dimension of embedding vectors.
The matrix is initialized with random values and updated during training.

\textbf{GRU Encoder} We design the encoder of VAE as a bi-directional GRU.
Sequentially, it deals with the input tokens, and propagates the upstream and downstream context through the hidden states, respectively in the forward and backward directions, as shown in~\cref{eq:encoder-forward} and~\cref{eq:encoder-backward}.
$emb$ maps a token to its embedding vector.
Finally, we sum up the last hidden states of both directions to get the final hidden state, as in~\cref{eq:encoder-sum}, and pass it to the next layer.
\begin{align*}
\overrightarrow{\textbf{h}_{i}} &= \overrightarrow{GRU}(emb(w_{i}), \overrightarrow{\textbf{h}_{i-1}}) \numberthis \label{eq:encoder-forward} \\
\overleftarrow{\textbf{h}_{i}} &= \overleftarrow{GRU}(emb(w_{i}), \overleftarrow{\textbf{h}_{i+1}}) \numberthis \label{eq:encoder-backward}\\
\textbf{h} &= \overrightarrow{\textbf{h}_{n}} + \overleftarrow{\textbf{h}_{n}} \numberthis \label{eq:encoder-sum}
\end{align*}

\textbf{Latent Variable} Based on the hidden state \textbf{h} from the encoder, we estimate the parameters of a Gaussian distribution with a fully-connected layer, which are the mean vector $\bm{\mu}$ and variance vector $\bm{\sigma}^2$. 
The latent variable $\textbf{z}$ is randomly sampled from this distribution. 
The equations are as follows:
\begin{align*}
\bm{\mu};\bm{\sigma^2} &= FC(\textbf{h})\\
\textbf{z} &= \bm{\mu} + \textbf{r} \cdot \textbf{e}^{\bm{\sigma^2} / 2}
\end{align*}
where $FC$ is a fully-connected layer and $\textbf{r}$ is a random vector from the standard normal distribution.

\textbf{GRU Decoder} The latent variable represents the key features of the original input in a highly abstract and compact way. 
The decoder works to reconstruct the input solely based on the latent variable. Iteratively, the decoder computes the hidden state $s_i$ at each step $i$ and reconstructs token $w_i'$, based on the previous state $s_{i-1}$ (or $\textbf{z}$ at step 0) and $w_{i-1}'$ generated in the previous step.
The equations are as follows:
\begin{align*}
\textbf{s}_{i} &= 
\begin{cases}
 GRU(emb(bos),\textbf{z})) & \text{i=0}\\
 GRU(emb(w_{i-1}'),\textbf{s}_{i-1})) & \text{i\textgreater{}0}
\end{cases} \\
p_i &= FC(\textbf{s}_i)\\
w_i' &= argmax(p_i) 
\end{align*}
where $bos$ is a special token indicating the start of a sentence, and $p_i \in \mathbb{R}^{o_w}$ represents the probability of $i$-th token to be generated.

\textbf{Loss} We measure the likelihood of reconstructing the original input with the Cross-Entropy (CE) loss. Hence, the ELBO loss introduced in \cref{sec:vae} can be computed as:
\begin{equation}
\mathcal{L} = - \frac{1}{n}\sum_{i=1}^{n}CELoss(w_i,w_i') +  KLDivergence(\bm{\mu},\bm{\sigma}^2) \nonumber
\end{equation} 
where $CELoss$ and $KLDivergence$ represents the calculation of the CE loss and KL divergence.

\subsubsection{The Filtering Algorithm}
\label{sec:trend}
We train the VAE model with a set of high-quality code search queries collected from near-real scenarios, which we call the \corpus.
After the training, the VAE model is able to recognize whether a query semantically resembles those in the corpus. 
We measure the reconstruction loss, i.e., the CE loss, of an input when fed to the VAE model, which just reflects how well it is within the training set distribution.
Intuitively, the loss value is the anomaly score gauging how far an input stays away from the queries in \corpus.
Comments with smaller losses are more likely to be query-appropriate.

\begin{figure}[t]
\centerline{\includegraphics[width=5cm]{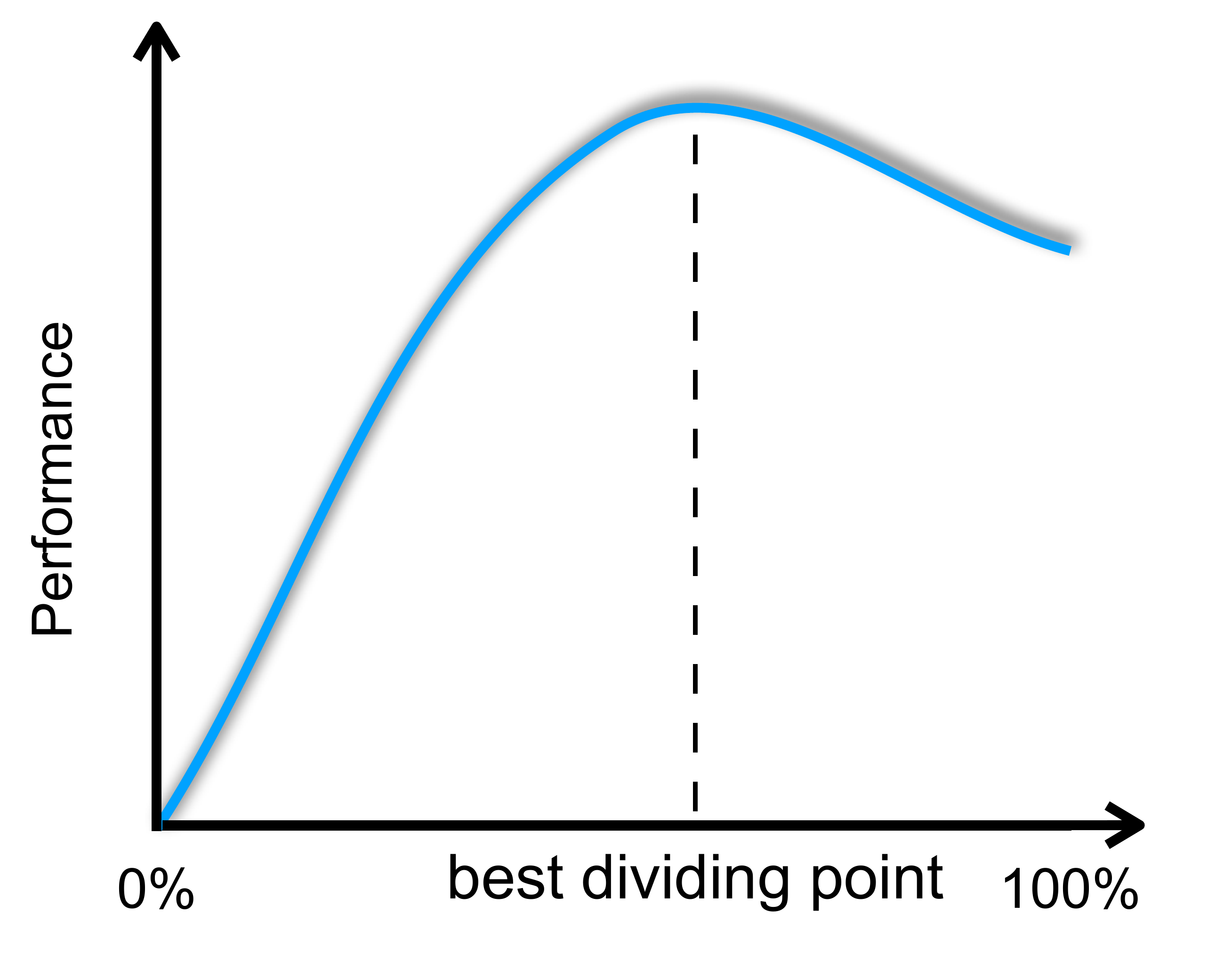}}
\caption{An illustration of the relation between the portion of retained data and the performance of the code search model trained with it.}
\label{fig:trend}
\end{figure}

To select comments resembling queries in \corpus, we sort the comments based on their reconstruction losses, in ascending order, and retain the top-ranked ones.
It is tricky to decide an appropriate dividing point for retaining the portion with better quality and discarding the remaining.
The less data we keep from the top, the higher the dataset quality. 
However, a sharp reduction in the data size hinders the performance of the trained code search model. 
\cref{fig:trend} shows a theoretical model illustrating the relation between the dividing point and the model performance.
As the amount of retained data increases, the model performance firstly increases and then decreases after reaching the peak. 
There is a trade-off between the quality and quantity of the dataset.

We leverage an unsupervised clustering algorithm, EM-GMM (Expectation-Maximization for Gaussian Mixture Model)~\cite{Dempster1977MaximumLF}, to decide the partition automatically. 
It is widely used to model the mixed distributions of a dataset.
For our task, EM-GMM divides a set of comments into the qualified and the unqualified groups based on the reconstruction loss. For each group, GMM fits a Gaussian probability density function and mixes them together as the distribution of the whole dataset, which can be represented as:
\begin{align*}
P(x) = \pi N(x|\mu_q,\sigma_q) + (1-\pi)N(x|\mu_{uq},\sigma_{uq}) 
\end{align*}
where $\pi$ is the mixture coefficient for the qualified group, ($\mu_q,\sigma_q$) and ($\mu_{uq},\sigma_{uq}$) are the parameters for the Gaussian probability density functions of the qualified and unqualified groups, respectively.
Finally, the EM algorithm~\cite{Dempster1977MaximumLF} is applied to estimate a set of optimal values for all the parameters.

Note that, to establish a high-quality code search dataset, all comments are processed together with their paired code snippets.
Hence, we obtain a set of comment-code pairs after applying this semantic filter, where the comments are syntactically and semantically close to natural user queries.

%% file: experiments.tex
\section{Experiment setup}

\begin{table*}[h]
\setlength\tabcolsep{2.8pt}

\caption{The Answered@k and MRR scores of the DeepCS and CARLCS models trained over different datasets.
}
\centering
\input{tables/rq1-results}
\label{tab:rq1_deepcs}
\end{table*}

\label{sec:experiment-setup}
We introduce the research questions, the basic experimental setup about the datasets and models, and the evaluation metrics used throughout the evaluation. The research questions we aim to answer include:

\noindent \textbf{RQ1:} How effective is our data cleaning framework?

\noindent \textbf{RQ2:} What is the impact of each filter component and each rule on the effectiveness of our framework?

\noindent \textbf{RQ3:} How is the dividing point determined by the clustering algorithm during the model-based filtering?

\subsection{Datasets}
\label{sec:dataset}
Three types of datasets are involved in our evaluation, including the training and validation datasets used to train and assess the performance of code search models and the \corpus used to develop our model-based filter.
To make the best use of existing resources, we focus on the Java programming language in this work, for which there have been the most public datasets and models in the field of neural code search.
Theoretically, our framework is language-independent and applicable to other programming languages with a proper adaptation of the filtering rules.

\subsubsection{Training Datasets}
We use the popular CodeSearchNet (CSN)~\cite{Husain2019CodeSearchNetCE} dataset to train all the code search models. 
CSN is a collection of datasets and benchmarks for semantic code retrieval. 
It extracts functions and their paired comments from Github repositories. 
It covers six programming languages, and we take the training dataset for Java, which contains 394,471 data points. 
We took the first sentence of each comment.
In what follows, we denote it as \textbf{CSN-t}.
Another widely used dataset in DeepCS~\cite{Gu2018DeepCS} is not included because the authors only released the processed data, but our framework cannot work without accessing the raw data.

\subsubsection{Validation Datasets}
We utilize human-annotated validation datasets to evaluate how well a code search model performs in a real-world scenario, and three widely used datasets are adopted.
It is noteworthy that the validation datasets are \emph{never} filtered by our framework in order to ensure the fairness of our experiments.
They are listed as follows:
\begin{itemize}[leftmargin=*]
\item \textbf{CSN-v} CSN also offers a validation benchmark for Java, containing query-code pairs collected from Bing and StackOverflow.
Human annotators are hired to rate the relevance between the query and the code.
Pairs with a score greater than $0$ are deemed as relevant, and there are 434 relevant pairs in total.
In the dataset, each pair is accompanied by 999 distractor code snippets.
It means, given a query, the code search model needs to retrieve the ground truth among 1000 candidates.

\item \textbf{CB} CosBench (CB)~\cite{Yan2020AreTC} is a validation dataset consisting of 52 selective queries from StackOverflow.
For each query, the authors prepared around ten paring code snippets as its ground truths, including its best answer on StackOverflow and several other matched code snippets selected from GitHub.
Additionally, there is a pool of 4,199,769 distractor code snippets.
The model needs to search the ground truths from a mixture with the complete code pool given a query.

\item \textbf{NCSED} Proposed in~\cite{Li2019NeuralCS}, the NCSED dataset contains 287 question queries manually collected from StackOverflow.
For each query, there are around three pairing code snippets selected from GitHub. 
The ground truths are mixed with other 4,716,814 distractor code snippets collected from GitHub.
The search model is required to retrieve the ground truths from the large corpus for a query.
\end{itemize}

The extremely
large search space in NCSED and CB makes it extremely hard for code search models to achieve a good performance, and the performances variations brought by data cleaning can also be too marginal to compare. Without loss of generality, for each query in NCSED and CB,
we construct 999 distractor snippets, following a similar fashion as CSN-v.

\subsubsection{Bootstrap Query Corpus} StackOverflow is an ideal source for collecting resemblers of actual user queries, though the quality of the pairing code snippets is hard to guarantee. 
It becomes an optimal choice to establish the \corpus.
We surveyed existing StackOverflow datasets in the code search field, and found that they were of severely limited size.
With their aim to collect high-quality question-code pairs, numerous questions were discarded due to the lack of qualified code answers.
Hence, to better facilitate the training of our VAE model, we determined to construct a question-only corpus from StackOverflow instead of using existing ones. 

According to a study~\cite{Nasehi2012WhatMA}, the StackOverflow questions can be divided into four types: ``Debug/Corrective'', ``Need-To-Know'', ``How-To-Do-It'' and ``Seeking-Different-Solution''. Among them, questions of the ``How-To-Do-It'' type are most relevant to queries for the code search task. 
Aiming to select the most qualified resemblers, we require the question titles to 1) start with ``how to'', 2) be tagged with ``Java'', and 3) pass the rule-based syntactic filter proposed in~\Cref{sec:rule} (except for the Interrogation rule).
In the end, 168,779 out of 1,709,703 Java-related question titles were retained.
Afterwards, we transformed them into declarative sentences by removing the starting ``how to'' and the question marks if any, thus forming the \corpus, which is used to train the VAE model later.

\subsection{Code Search Model}
Two code search models, DeepCS~\cite{Gu2018DeepCS} and CARLCS~\cite{Shuai2020ImprovingCS}, are used in our experiments. 
They are designed with representative architectures among most neural code search models.
DeepCS is based on the Siamese architecture, and CARLCS is an Interaction-based network~\cite{mitra2018introduction}. 
The Siamese architecture consists of two DL models to represent the query and code, respectively, with independent embedding vectors, and the similarity between these vectors is used to measure the relevance between query and code.
The Interaction-based network compares the query and code directly by generating an interaction matrix to reflect their relevance.

When training the models with our training dataset, we adopted the recommended settings for all the hyper-parameters, except for the training epoch of the DeepCS model.
In order to save some time and computation resources, we set the maximum training epoch of DeepCS to 100 instead of the recommended 500.
Without loss of fairness, the same setting has been used for training with the dataset either before or after the data cleaning.
This change should not affect the evaluation conclusion on the effectiveness of our framework, which focuses more on whether the model performance improves after removing the noises instead of its absolute level.

\subsection{Evaluation Metrics}
Two widely used metrics are adopted in our experiments to evaluate the code retrieval performance.
\begin{itemize}[leftmargin=*]
\item \textbf{Answered@k}: Answered@k (abbrev. A@k) is the number of queries answered by snippets in the top-k results.

\item \textbf{Mean Reciprocal Rank (MRR)}: MRR is the average of the reciprocal ranks of the ground truth in the result list.
\end{itemize}

\section{Results}
\label{sec:results}
In this section, we show the experimental results and answer the research questions. 
Measures for both evaluation metrics are reported as the medium over five independent runs.

\subsection{RQ1: Effectiveness}
\label{sec:rq1}

This experiment evaluates the effectiveness of our data cleaning solution as a pre-processing step when training neural code search models.
Specifically, one training dataset (CSN-t), two code search models (DeepCS and CARLCS), and three validation datasets (CSN-v, CB, and NCSED) are used in the evaluation.
Thus, we have six $(1 \times 2 \times 3)$ experimental settings in total.
During experiments, a relatively smaller filtered training set will be derived from CSN-t after our framework is applied for the data cleaning.
To also benchmark the performance variation brought by the size shrinking, we further derive a controlled training set by randomly selecting from CSN-t an equivalent number of data as the filtered set.
We observe the model performance resulted from training with these three datasets respectively.

The model performance is measured with four evaluation metrics, namely, A@1, A@5, A@10, and MRR, and the results are shown in~\Cref{tab:rq1_deepcs}.
Under all the six experimental settings, our data cleaning framework demonstrates a positive influence on the model's searching ability and helps it hit the best score.
On average of the three validation datasets, DeepCS trained over the filtered data outperforms the one trained over original data by 21.3\% A@1, 17.4\% A@5, 7.8\% A@10, and 19.2\% MRR.
Correspondingly, the improvements of CARLCS are 63.3\% A@1, 58.6\% A@5, 16.0\% A@10, and 38.6\% MRR. 
Regarding the MRR on the three validation datasets, CSN-v,CB, and NCSED, DeepCS achieves 0.512, 0.644 and 0.271, and CARLCS achieves 0.302, 0.056 and 0.168, respectively.
Basically, DeepCS and CARLCS are boosted to their new best records, and CARLCS sees a greater improvement.
Note that the A@1 score of CARLCS over NCSED is increased by 75.0\% (from 20 to 35), which is an extraordinary improvement.

Overall, with around half of the data quantity and half of the training time, models trained over the filtered data achieve a significant improvement on the number of answered queries and the rank of ground truth in search results.

\begin{tcolorbox}[size=title]
{\textbf{Answer to RQ1:}}
Our filtering framework produces a high-quality query-code dataset, which shortens the training time by reducing the training data and effectively improves the performance of the code search model under a real-world application scenario.
\end{tcolorbox}

\begin{table*}[t]
\caption{Results of the ablation experiments on the filter components.
}
{
\setlength\tabcolsep{2.8pt}
\begin{center}
    \input{tables/merged-rq2-results}
\end{center}
}
\label{tab:rq2}
\end{table*}

\subsection{RQ2: The impact of each filter component and each rule}
\label{sec:re2}

We evaluate the effectiveness of each filter component with ablation experiments and conduct manual inspection on the queries accepted/rejected by each syntactic rule and the model-based filter to study their precision in identifying noises. 

Each time, one of the two filter components is muted for the ablation experiments. 
We observe the model performance after training with such derived filtered dataset and compare it with their previous performance (in \Cref{sec:rq1}).
If the performance declines compared with when both filters are enabled, we can infer a positive impact of the muted component on the framework effectiveness. 
We evaluate the performance of DeepCS and CARLCS trained under ablation and report the results in~\Cref{tab:rq2}. 
The removal of any filter leads to worse performance scores. 
Without the model-based filter, the A@1, A@5, A@10, and MRR scores of DeepCS on the three validation sets reduce by 5.4\%, 5.6\%, 3.6\%, and 4.6\% on average. CARLCS sees a much more severe deduction, and on average, A@1, A@5, A@10, and MRR decrease by 35.5\%, 18.7\%, 10.4\%, and 20.6\%. 
After removing the rule-based filter, the performance of DeepCS averagely drops by 8.4\% A@1, 6.4\% A@5, 5.2\% A@10, and 7.6\% MRR. Meanwhile, the average reduction percentages of CARLCS on all the validation sets are 32.7\% A@1, 15.5\% A@5, 9.9\% A@10, and 14.3\% MRR. 
It is noteworthy that the A@1 score of CARLCS on NCSED drops from 35 to 18 when the rule filter is muted, indicating that the ruleset plays a very influential part during the data cleaning.

For the manual inspection, two annotators, with over two years' development experience, are hired to rate how likely a sentence is to be used as a code search query.
The rating score ranges from 0 to 2, where 0 means worst and 2 best.
There are 11 groups of data to annotate, including eight groups of comments rejected by each rule, the group of comments discarded by the model filer, the original CSN-t dataset, and the filtered dataset after the two-filter cleaning.
The last two groups are for comparison purposes.
For rules focusing on detachable features, i.e., the Parentheses and HTML tags, we let the annotators judge how well the removed part can help with a query expression.
We sample a subset of data from its full set for each group. 
The sample size $ss$ of each group is computed by a statistical formula which is extracted from \cite{cochran1977sampling},
$
ss = \frac{z^2*p*(1-p)/c^2}{1+\frac{z^2*p*(1-p)/(c^2-1)}{population}},
$
where $population$ is the size of the entire dataset, $p$ is the standard deviation of the population, $c$ is the confidence interval (margin of error), $z$ is the Z-Score determined by the confidence level. In this experiment, we choose to work with a 95\% confidence level (i.e., 1.96 Z-Score according to \cite{israel1992determining}), a standard deviation of 0.5 (0.5 is the maximum standard deviation, and a safe choice given the exact figure unknown), and a confidence interval of 5\%.
We also measure the agreement between the two annotators with Cohen's Kappa~\cite{Cohen1960A}, which is 0.69 and within the range of fair to good.

For each data, we finalize its score as the average of scores from the two annotators and display the statistics in \Cref{tab:rq4}.
We report the number of data examined in each group, the respective portion of data scored as 0 or no less than 1, and the group's average score, in the last four columns.
In general, the comments rejected by either the rule-based filter or the model-based filter poorly resemble real user queries, with 96.9\% and 85.9\% of them receiving a score of 0 and the average scores being as low as 0.04 and 0.20, respectively.
Still, it comes at an acceptable cost of losing a small set of good quality data, where 3.1\% and 14.1\% of the discarded data by the two filters score at least 1.
Each of the eight rules rejects code comments in an effective way, with four of them rejecting non-query-like data at 100\% precision.
The precision of the Parentheses rule is relatively low, where 11\% of the discarded data is of high quality. 
In the future, when deciding whether the content inside the parentheses should be removed, a more refined rule can be derived. 
Also, the model-based semantic filter is accompanied by a larger sacrifice, indicating it as a more challenging task.

\begin{table}[t]
\caption{Results of the manual inspection.}

{
\setlength\tabcolsep{1.7pt}
\begin{center}
    \input{tables/manual-evaluation}
\end{center}
}
\label{tab:rq4}
\end{table}

Overall, through the two-phase filtering, the average likeness score increases from 0.27 to 0.61. 
In particular, the portion of non-query-like data drops from 79.0\% to 59.2\%, and the portion of highly query-like data scoring at least 1 improves from 21\% to 40.8\%.
There are still many comments inappropriate to be seen as code search queries, but our data cleaning framework makes a substantial contribution to alleviating the situation.
We call for more attention to be drawn to overcoming related challenges.

\begin{tcolorbox}[size=title]
{\textbf{Answer to RQ2:}} 
Each filter and rule in our framework demonstrates a positive contribution to the effectiveness.
The full setting boosts it to the best performance. 
However, there remain many unqualified comments even after the filtering, and it calls for more attention to be paid from the community.
\end{tcolorbox}

\begin{table*}[t]
\caption{Results of changing the EM-GMM to other methods.}
\setlength\tabcolsep{2.3pt}
\begin{center}
    \input{tables/cluster-results}
\end{center}
\label{tab:rq3}
\end{table*}

\subsection{RQ3: Quality of dividing point determined in the  model-based filtering}

In the model-based filter, we use EM-GMM to decide the dividing point between the qualified and the unqualified groups.
To assess the quality of the dividing point, we observe the model performance resulting from alternative dividing points, including fix proportions and the one decided by K-means, another widely used clustering algorithm.
For the fixed proportions, we set a 25\% step and select 25\%, 50\%, 75\%, and 100\% top-ranked comments, respectively.

The results on DeepCS and CARLCS are reported in Table~\ref{tab:rq3}. 
For DeepCS, EM-GMM outperforms K-means and the fixed proportions on all the validation sets. Compared with the second-best partition, 75\%, EM-GMM still achieves higher average performances by 4.3\% A@1, 4.2\% A@5, 4.3\% A@10, and 3.3\% MRR. The superiority is also observed on CARLCS at every metric, and EM-GMM outperforms K-means on average of CSN-v and NCSED by 30.4\% A@1, 5.9\% A@5, 2.8\% A@10, and 15.0\% MRR.

EM-GMM ultimately retains 192,031 data points, accounting for 67.3\% of the original dataset,
which locates between 57.5\%, the dividing point set by K-means and 75\%.
As discussed in~\Cref{sec:trend}, the relation between the data quantity and the model performance should be a convex function.
According to the property of the convex function, if there exists another optimal dividing point, it would locate between 57.5\% and 75.0\%.
Therefore, EM-GMM successfully identifies an optimal solution of the dividing point with an error less than 9.8\% (calculated by $67.3\%-57.5\%$).

\begin{tcolorbox}[size=title]
{\textbf{Answer to RQ3:}} EM-GMM produces a better approximation of the best dividing point for the datasets and is adequate to be used in the framework.
\end{tcolorbox}

%% file: tables/rq1-results.tex
\begin{tabular}{|c|cc|lc|c|cl|cl|cl|cl|} 
\hline
\textbf{Model} & \textbf{Test~Set} & \textbf{\#Query} & \multicolumn{1}{c}{\textbf{Train~Set}} & \textbf{\#Pairs} & \textbf{Train~Hours} & \multicolumn{2}{c|}{\textbf{A@1}} & \multicolumn{2}{c|}{\textbf{A@5}} & \multicolumn{2}{c|}{\textbf{A@10}} & \multicolumn{2}{c|}{\textbf{MRR}} \\ 
\hline
\multirow{9}{*}{DeepCS} & \multirow{3}{*}{CSN-v} & \multirow{3}{*}{434} & CSN-t~(all) & 394,471 & 8~h & 123 &  & 248 &  & 306 &  & 0.407~ &  \\
 &  &  & CSN-t~(controlled) & 192,031 & 4~h & 107 &  & 235 &  & 275 &  & 0.376~ &  \\
 &  &  & CSN-t~(filtered) & 192,031 & 4~h & \textbf{168} & 36.6$\%\uparrow$ & \textbf{299} & 20.6$\%\uparrow$ & \textbf{348} & 13.7$\%\uparrow$ & \textbf{0.512}~ & 26.0$\%\uparrow$ \\ 
\cline{2-14}
 & \multirow{3}{*}{CB} & \multirow{3}{*}{52} & CSN-t~(all) & 394,471 & 8~h & 25 &  & 31 &  & 38 &  & 0.522~ &  \\
 &  &  & CSN-t~(controlled) & 192,031 & 4~h & 20 &  & 26 &  & 28 &  & 0.438~ &  \\
 &  &  & CSN-t~(filtered) & 192,031 & 4~h & \textbf{29} & 16.0$\%\uparrow$ & \textbf{38} & 22.6$\%\uparrow$ & \textbf{40} & 5.3$\%\uparrow$ & \textbf{0.644}~ & 23.3$\%\uparrow$ \\ 
\cline{2-14}
 & \multirow{3}{*}{NCSED} & \multirow{3}{*}{287} & CSN-t~(all) & 394,471 & 8~h & 44 &  & 101 &  & 136 &  & 0.250~ &  \\
 &  &  & CSN-t~(controlled) & 192,031 & 4~h & 31 &  & 90 &  & 135 &  & 0.210~ &  \\
 &  &  & CSN-t~(filtered) & 192,031 & 4~h & \textbf{49} & 11.4$\%\uparrow$ & \textbf{110} & 8.9$\%\uparrow$ & \textbf{142} & 4.4$\%\uparrow$ & \textbf{0.271~} & 8.4$\%\uparrow$ \\ 
\hline
\multirow{9}{*}{CARLCS} & \multirow{3}{*}{CSN-v} & \multirow{3}{*}{434} & CSN-t~(all) & 394,471 & 6~h & 54 &  & 210 &  & 292 &  & 0.283~ &  \\
 &  &  & CSN-t~(controlled) & 192,031 & 3~h & 54 &  & 202 &  & 284 &  & 0.281~ &  \\
 &  &  & CSN-t~(filtered) & 192,031 & 3~h & \textbf{62} & 14.8$\%\uparrow$ & \textbf{221} & 5.2$\%\uparrow$ & \textbf{296} & 1.4$\%\uparrow$ & \textbf{0.302}~ & 6.7$\%\uparrow$ \\ 
\cline{2-14}
 & \multirow{3}{*}{CB} & \multirow{3}{*}{52} & CSN-t~(all) & 394,471 & 6~h & 1 &  & 2 &  & 6 &  & 0.038~ &  \\
 &  &  & CSN-t~(controlled) & 192,031 & 3~h & 0 &  & 1 &  & 3 &  & 0.012~ &  \\
 &  &  & CSN-t~(filtered) & 192,031 & 3~h & \textbf{2} & 100.0$\%\uparrow$ & \textbf{4} & 100.0$\%\uparrow$ & \textbf{7} & 16.7$\%\uparrow$ & \textbf{0.056}~ & 49.4$\%\uparrow$ \\ 
\cline{2-14}
 & \multirow{3}{*}{NCSED} & \multirow{3}{*}{287} & CSN-t~(all) & 394,471 & 6~h & 20 &  & 34 &  & 57 &  & 0.105~ &  \\
 &  &  & CSN-t~(controlled) & 192,031 & 3~h & 15 &  & 33 &  & 49 &  & 0.097~ &  \\
 &  &  & CSN-t~(filtered) & 192,031 & 3~h & \textbf{35} & 75.0$\%\uparrow$ & \textbf{58} & 70.6$\%\uparrow$ & \textbf{74} & 29.8$\%\uparrow$ & \textbf{0.168}~ & 59.9$\%\uparrow$ \\
\hline
\end{tabular}

%% file: tables/merged-rq2-results.tex
\begin{tabular}{|c|cc|lc|cc|cc|cc|cc|} 
\hline
\textbf{Model} & \textbf{Test Set} & \textbf{\#Query} & \multicolumn{1}{c}{\textbf{Train Set}} & \textbf{\#Pairs} & \multicolumn{2}{c|}{\textbf{A@1}} & \multicolumn{2}{c|}{\textbf{A@5}} & \multicolumn{2}{c|}{\textbf{A@10}} & \multicolumn{2}{c|}{\textbf{MRR}} \\ 
\hline
\multirow{12}{*}{DeepCS} & \multirow{4}{*}{CSN-v} & \multirow{4}{*}{434} & CSN-t (all) & 394,471 & 123 &  & 248 &  & 306 &  & 0.407~ &  \\
 &  &  & CSN-t (filtered) & 192,031 & \textbf{168} &  & \textbf{299} &  & \textbf{348} &  & \textbf{0.512~} &  \\
 &  &  & Rule Filter only & 285,372 & 157 & 6.5$\%\downarrow$ & 283 & 5.4$\%\downarrow$ & 335 & 3.7$\%\downarrow$ & 0.490~ & 4.3$\%\downarrow$ \\
 &  &  & Model filter only & 286,306 & 158 & 6.0$\%\downarrow$ & 276 & 7.7$\%\downarrow$ & 323 & 7.2$\%\downarrow$ & 0.491~ & 4.2$\%\downarrow$ \\ 
\cline{2-13}
  & \multirow{4}{*}{CB} & \multirow{4}{*}{52} & CSN-t (all) & 394,471 & 25 &  & 31 &  & 38 &  & 0.522~ &  \\
 &  &  & CSN-t (filtered) & 192,031 & \textbf{29} &  & \textbf{38} &  & \textbf{40} &  & \textbf{0.644~} &  \\
 &  &  & Rule Filter only & 285,372 & 28 & 3.4$\%\downarrow$ & 35 & 7.9$\%\downarrow$ & 38 & 5.0$\%\downarrow$ & 0.598~ & 7.2$\%\downarrow$ \\
 &  &  & Model filter only & 286,306 & 24 & 17.2$\%\downarrow$ & 35 & 7.9$\%\downarrow$ & 38 & 5.0$\%\downarrow$ & 0.539~ & 16.3$\%\downarrow$ \\ 
\cline{2-13}
& \multirow{4}{*}{NCSED} & \multirow{4}{*}{287} & CSN-t (all) & 394,471 & 44 &  & 101 &  & 136 &  & 0.250~ &  \\
 &  &  & CSN-t (filtered) & 192,031 & \textbf{49} &  & \textbf{110} &  & \textbf{142} &  & \textbf{0.271~} &  \\
 &  &  & Rule Filter only & 285,372 & 46 & 6.1$\%\downarrow$ & 106 & 3.6$\%\downarrow$ & 139 & 2.1$\%\downarrow$ & 0.265~ & 2.3$\%\downarrow$ \\
 &  &  & Model filter only & 286,306 & 48 & 2.0$\%\downarrow$ & 106 & 3.6$\%\downarrow$ & 137 & 3.5$\%\downarrow$ & 0.264~ & 2.4$\%\downarrow$ \\ 
\hline
\multirow{12}{*}{CARLCS} & \multirow{4}{*}{CSN-v} & \multirow{4}{*}{434} & CSN-t (all) & 394,471 & 54 &  & 210 &  & 292 &  & 0.283~ &  \\
 &  &  & CSN-t (filtered) & 192,031 & \textbf{62} &  & \textbf{221} &  & \textbf{296} &  & \textbf{0.302~} &  \\
 &  &  & Rule Filter only & 285,372 & 57 & 8.1$\%\downarrow$ & 211 & 4.5$\%\downarrow$ & 288 & 2.7$\%\downarrow$ & 0.293~ & 3.0$\%\downarrow$ \\
 &  &  & Model filter only & 286,306 & 57 & 8.1$\%\downarrow$ & 219 & 0.9$\%\downarrow$ & 294 & 0.7$\%\downarrow$ & 0.300~ & 0.6$\%\downarrow$ \\ 
\cline{2-13}
& \multirow{4}{*}{CB} & \multirow{4}{*}{52} & CSN-t (all) & 394,471 & 1 &  & 2 &  & 6 &  & 0.038~ &  \\
 &  &  & CSN-t (filtered) & 192,031 & \textbf{2} &  & \textbf{4} &  & \textbf{7} &  & \textbf{0.056~} &  \\
 &  &  & Rule Filter only & 285,372 & 1 & 50.0$\%\downarrow$ & 2 & 50.0$\%\downarrow$ & 5 & 28.6$\%\downarrow$ & 0.039~ & 31.2$\%\downarrow$ \\
 &  &  & Model filter only & 286,306 & 1 & 50.0$\%\downarrow$ & 3 & 25.0$\%\downarrow$ & 6 & 14.3$\%\downarrow$ & 0.049~ & 14.0$\%\downarrow$ \\
\cline{2-13}
  & \multirow{4}{*}{NCSED} & \multirow{4}{*}{287} & CSN-t (all) & 394,471 & 20 &  & 34 &  & 57 &  & 0.105~ &  \\
 &  &  & CSN-t (filtered) & 192,031 & \textbf{35} &  & \textbf{58} &  & \textbf{74} &  & \textbf{0.168~} &  \\
 &  &  & Rule Filter only & 285,372 & 18 & 48.6$\%\downarrow$ & 57 & 1.7$\%\downarrow$ & \textbf{74} & 0.0$\%\downarrow$ & 0.122~ & 27.6$\%\downarrow$ \\
 &  &  & Model filter only & 286,306 & 21 & 40.0$\%\downarrow$ & 46 & 20.7$\%\downarrow$ & 63 & 14.9$\%\downarrow$ & 0.120~ & 28.4$\%\downarrow$ \\ 
\hline
\end{tabular}

%% file: tables/manual-evaluation.tex
\begin{tabular}{|c|c|c|ccc|}
\hline
\multirow{2}{*}{\textbf{Type}} & \multirow{2}{*}{\textbf{Rule}} & \multirow{2}{*}{\textbf{\#}} & \multicolumn{3}{c|}{\textbf{Likeness score}} \\ \cline{4-6} 
 &  &  & \textbf{$=$0} & \textbf{$\geq$1} & \textbf{Avg.} \\ \hline
Origin & - & 394,471 & 79.0\% & 21.0\% & 0.27 \\ \hline
\multirow{9}{*}{\begin{tabular}[c]{@{}c@{}}Discarded by\\ Rule Filer\end{tabular}} & HTML Tag & 32,989 & 100.0\% & 0.0\% & 0.00 \\
 & Parentheses & 19,305 & 89.0\% & 11.0\% & 0.15 \\
 & Javadoc Tags & 47,106 & 94.9\% & 5.1\% & 0.08 \\
 & URLs & 640 & 97.3\% & 2.7\% & 0.05 \\
 & Non-Eng. Lan. & 6,503 & 100.0\% & 0.0\% & 0.00 \\
 & Punctuation & 39,032 & 100.0\% & 0.0\% & 0.00 \\
 & Interrogation & 516 & 100.0\% & 0.0\% & 0.00 \\
 & Short Sentence & 15,186 & 97.3\% & 2.7\% & 0.04 \\ \cline{2-6} 
 & In total & 161,277 & 96.9\% & 3.1\% & 0.04 \\ \hline
\begin{tabular}[c]{@{}c@{}}Discarded by \\ Model Filer\end{tabular} & - & 93,457 & 85.9\% & 14.1\% & 0.20 \\ \hline
Retained & - & 192,031 & 59.2\% & 40.8\% & \textbf{0.61} \\ \hline
\end{tabular}

%% file: tables/cluster-results.tex
\begin{tabular}{|c|l|c|cccc|cccc|cccc|} 
\hline
\multirow{2}{*}{\textbf{Model}} & \multicolumn{1}{c|}{\multirow{2}{*}{\textbf{Dividing~Point}}} & \multirow{2}{*}{\textbf{\#}} & \multicolumn{4}{c|}{\textbf{CSN-v~(434)}} & \multicolumn{4}{c|}{\textbf{CB~(52)}} & \multicolumn{4}{c|}{\textbf{NCSED~(287)}} \\ 
\cline{4-15}
 & \multicolumn{1}{c|}{} &  & \textbf{A@1} & \textbf{A@5} & \textbf{A@10} & \textbf{MRR} & \textbf{A@1} & \textbf{A@5} & \textbf{A@10} & \textbf{MRR} & \textbf{A@1} & \textbf{A@5} & \textbf{A@10} & \textbf{MRR} \\ 
\hline
\multirow{6}{*}{DeepCS} & Percentile~(25\%) & 71,343 & 134 & 240 & 285 & 0.420~ & 22 & 31 & 34 & 0.500~ & 40 & 80 & 115 & 0.215~ \\
 & Percentile~(50\%) & 142,686 & 138 & 277 & 324 & 0.459~ & 23 & 33 & 36 & 0.528~ & 34 & 91 & 129 & 0.212~ \\
 & KMeans~(57.5\%) & 164,194 & 146 & 274 & 321 & 0.471~ & 23 & 34 & 39 & 0.530~ & 39 & 99 & 127 & 0.229~ \\
 & EM-GMM~(67.3\%) & 192,031 & \textbf{168} & \textbf{299} & \textbf{348} & \textbf{0.512~} & \textbf{29} & \textbf{38} & \textbf{40} & \textbf{0.644~} & \textbf{49} & \textbf{110} & \textbf{142} & \textbf{0.271~} \\
 & Percentile~(75\%) & 214,029 & 160 & 284 & 332 & 0.505~ & 28 & 36 & 38 & 0.604~ & 47 & 108 & 138 & 0.266~ \\
 & Percentile~(100\%) & 285,372 & 157 & 283 & 335 & 0.490~ & 28 & 35 & 38 & 0.598~ & 46 & 106 & 139 & 0.265~ \\ 
\hline
\multirow{6}{*}{CARLCS} & Percentile~(25\%) & 71,343 & 51 & 190 & 270 & 0.264~ & 0 & 2 & 6 & 0.031~ & 6 & 23 & 35 & 0.053~ \\
 & Percentile~(50\%) & 142,686 & 57 & 217 & 290 & 0.295~ & 1 & 2 & 5 & 0.031~ & 8 & 32 & 48 & 0.070~ \\
 & KMeans~(57.5\%) & 164,194 & 58 & 220 & 292 & 0.299~ & 1 & 2 & 5 & 0.032~ & 24 & 44 & 59 & 0.119~ \\
 & EM-GMM~(67.3\%) & 192,031 & \textbf{62} & \textbf{221} & \textbf{296} & \textbf{0.302~} & \textbf{2} & \textbf{4} & \textbf{7} & \textbf{0.056~} & \textbf{35} & \textbf{58} & \textbf{74} & \textbf{0.168~} \\
 & Percentile~(75\%) & 214,029 & 61 & 216 & 288 & 0.298~ & 1 & 3 & 5 & 0.035~ & 22 & 53 & 72 & 0.130~ \\
 & Percentile~(100\%) & 285,372 & 57 & 211 & 288 & 0.293~ & 1 & 2 & 5 & 0.039~ & 18 & 57 & 74 & 0.122~ \\
\hline
\end{tabular}

%% file: application.tex
\section{Application}

\label{sec:application}
This section presents the applications of our filtering framework, including a proof-of-concept data cleaning toolbox and a high-quality code search dataset.

\subsection{\tool: Natural Language Query Filter}
We release the implementation of our filtering framework as a third-party Python library, \underline{N}atural \underline{L}anguage \underline{Q}uery \underline{F}ilter (\tool), which is designed to systemically filter queries for neural code search models. 
As a lightweight library with convenient  APIs, \tool can be easily integrated into the development pipeline of any code search model. Besides, \tool is extensible at several features to ensure its applicability in a wide range of contexts:

\noindent\textbf{Extensible Ruleset} The ruleset in \tool is configurable, which enables users to specify the rules based on the characteristics of their own data. Besides, \tool accepts user-defined functions as a part of rule-based filtering. One can easily extend the filter implementation by creating the filtering function for any new rule.

\noindent\textbf{Open-source Filtering Model} \tool requires a trained VAE model in the model-based filter. We release the source code for training the VAE model used in this paper. Following the instructions, users can easily train a new model with their own \corpus, which may boost the filtering performance further.

\noindent\textbf{Tunable Dividing Proportion} Besides the recommended clustering method, EM-GMM, \tool also provides an interface accepting user-defined dividing points. Users can create their own method for finding the dividing point and configure \tool to adopt it easily.

\subsection{\dataset: Codebase Paired with Filtered Comments}

We build and release a \underline{Co}debase paired with \underline{Fi}ltered \underline{C}omments (\dataset) for Java programming language.

\subsubsection{Dataset Building}
We collect the source code of Java repositories from Github according to the list maintained by Libraries.io~\cite{Nesbitt2017LibrariesioOS}, 
From these files, we extract the methods and corresponding comments using the scripts provided by CodeSearchNet~\cite{Husain2019CodeSearchNetCE}. 
In the end, 2,475,692 raw comment-code pairs are obtained. 
Through the processing with \tool, there are 1,048,519 data points left in the cleaned query-code dataset. 
Detailed statistics of the dataset during filtering are reported in~\Cref{tab:change}.

\begin{table}[t]
\caption{The statistics during the data filtering.}
\input{tables/change}
\label{tab:change}
\end{table}

\begin{table}[t]
\caption{A comparison between the training datasets for code search tasks.}

\begin{center}
\input{tables/comparison}
\end{center}
\label{tab:comparison}
\end{table}

\subsubsection{Dataset Comparison}
We compare \dataset, on the query quality, with several other datasets currently used in neural code search research.
Following the same manual inspection convention as in~\cref{sec:re2}, the annotators rate the queries sampled from each dataset, reported in~\cref{tab:comparison}. Again, we measure the agreement level between the two annotators with Cohen's Kappa, which is 0.73 and within the range of fair to good. 
Among all the datasets collected from Github, \dataset receives the highest score on data quality, but there is still a gap compared with the StackOverflow dataset, StaQC. 
Indeed, the datasets collected from StackOverflow have high-quality queries, but they suffer from the unstable code quality in answers\cite{Terragni2016CSNIPPEXAS,Zhang2018AreCE}.
With our filtering framework, a Github dataset with better quality is established.

Besides the user study, we also experimentally compare \dataset with CSN-t. We train the DeepCS and CARLCS models with three datasets: CSN-t, \dataset, and a controlled \dataset (same size as CSN-t). 
The model trained with \dataset outperforms other experimental settings on the three validation datasets (CSV-v, CB, and NCSED). 
The detailed results are reported in \Cref{tab:cofic}.

\begin{table*}[t]
\setlength\tabcolsep{2.3pt}
\caption{The results of the experimental comparison between COFIC and CSN-t.}
\begin{center}
\input{tables/cofic}
\end{center}
\label{tab:cofic}
\end{table*}

%% file: tables/change.tex
\begin{tabular}{|c|ccc|}
\hline
Step  & Rule & \#Discarded & \#Retained \\ \hline
\multirow{8}{*}{Rule-based} & HTML tags& 189,250  & 2,475,692 \\
  & Parentheses & 129,130  & 2,475,692\\
  & Javadoc tags & 423,313 & 2,052,379 \\
  & URLs  & 3,119 & 2,049,260 \\
  & Non-English Languages & 67,943  & 1,981,317 \\
  & Punctuation & 201,881  & 1,779,436 \\
  & Interrogation & 3,300 & 1,776,136 \\
  & Short Sentence & 112,133  & 1,664,003\\ \hline
Model-based   & -& 615,484 & \textbf{1,048,519}  \\ \hline
\end{tabular}

%% file: tables/comparison.tex
\begin{tabular}{|ccc|cc|}
\hline
Dataset & Source & Language & Likeness & \# \\ \hline
COFIC & Github & Java & 0.52 & 1 M \\
CSN (Java)\cite{Husain2019CodeSearchNetCE} & Github & Java & 0.27 & 543 K \\
Hu et al.\cite{Hu2018DeepCC} & Github & Java & 0.48 & 69 K \\
\multirow{2}{*}{StaQC\cite{Yao2018StaQCAS}} & \multirow{2}{*}{StackOverflow} & Python & 0.80 & 148 K \\
 &  & SQL & 0.80 & 120 K \\
Barone et al.\cite{Barone2017APC} & Github & Python & 0.43 & 150 K \\ \hline
\end{tabular}

%% file: tables/cofic.tex
\begin{tabular}{|c|cc|lc|cc|cc|cc|cc|} 
\hline
\textbf{Model} & \textbf{Test Set} & \textbf{\#Query} & \multicolumn{1}{c}{\textbf{Train Set}} & \textbf{\#Pairs} & \multicolumn{2}{c|}{\textbf{A@1}} & \multicolumn{2}{c|}{\textbf{A@5}} & \multicolumn{2}{c|}{\textbf{A@10}} & \multicolumn{2}{c|}{\textbf{MRR}} \\ 
\hline
\multirow{9}{*}{DeepCS} & \multirow{3}{*}{CSN-v} & \multirow{3}{*}{434} & CSN-t~ & 394,471 & 123 &  & 248 &  & 306 &  & 0.407~ &  \\
 &  &  & COFIC (controlled) & 394,471 & 188 & 52.8$\%\uparrow$ & 297 & 19.8$\%\uparrow$ & 344 & 12.4$\%\uparrow$ & 0.555~ & 36.3$\%\uparrow$ \\
 &  &  & COFIC & 1,048,519 & \textbf{191} & 55.3$\%\uparrow$ & \textbf{314} & 26.6$\%\uparrow$ & \textbf{354} & 15.7$\%\uparrow$ & \textbf{0.577~} & 41.9$\%\uparrow$ \\ 
\cline{2-13}
 & \multirow{3}{*}{NCSED} & \multirow{3}{*}{287} & CSN-t~ & 394,471 & 44 &  & 101 &  & 136 &  & 0.250~ &  \\
 &  &  & COFIC (controlled) & 394,471 & 58 & 31.8$\%\uparrow$ & 109 & 7.9$\%\uparrow$ & 137 & 0.7$\%\uparrow$ & 0.281~ & 12.4$\%\uparrow$ \\
 &  &  & COFIC & 1,048,519 & \textbf{72} & 63.6$\%\uparrow$ & \textbf{118} & 16.8$\%\uparrow$ & \textbf{148} & 8.8$\%\uparrow$ & \textbf{0.327~} & 31.0$\%\uparrow$ \\ 
\cline{2-13}
 & \multirow{3}{*}{CB} & \multirow{3}{*}{52} & CSN-t~ & 394,471 & 25 &  & 31 &  & 38 &  & 0.522~ &  \\
 &  &  & COFIC (controlled) & 394,471 & 34 & 36.0$\%\uparrow$ & 38 & 22.6$\%\uparrow$ & 39 & 2.6$\%\uparrow$ & 0.693~ & 32.6$\%\uparrow$ \\
 &  &  & COFIC & 1,048,519 & \textbf{38} & 52.0$\%\uparrow$ & \textbf{39} & 25.8$\%\uparrow$ & \textbf{41} & 7.9$\%\uparrow$ & \textbf{0.744~} & 42.3$\%\uparrow$ \\ 
\hline
\multirow{9}{*}{CARLCS} & \multirow{3}{*}{CSN-v} & \multirow{3}{*}{434} & CSN-t~ & 394,471 & 54 &  & 210 &  & 292 &  & 0.283~ &  \\
 &  &  & COFIC (controlled) & 394,471 & 66 & 22.2$\%\uparrow$ & 237 & 12.9$\%\uparrow$ & 319 & 9.2$\%\uparrow$ & 0.328~ & 15.8$\%\uparrow$ \\
 &  &  & COFIC & 1,048,519 & \textbf{69} & 27.8$\%\uparrow$ & 247 & 17.6$\%\uparrow$ & \textbf{322} & 10.3$\%\uparrow$ & \textbf{0.339~} & 19.8$\%\uparrow$ \\ 
\cline{2-13}
 & \multirow{3}{*}{NCSED} & \multirow{3}{*}{287} & CSN-t~ & 394,471 & 20 &  & 34 &  & 57 &  & 0.105~ &  \\
 &  &  & COFIC (controlled) & 394,471 & 34 & 70.0$\%\uparrow$ & 68 & 100.0$\%\uparrow$ & 87 & 52.6$\%\uparrow$ & 0.172~ & 64.1$\%\uparrow$ \\
 &  &  & COFIC & 1,048,519 & \textbf{47} & 135.0$\%\uparrow$ & \textbf{75} & 120.6$\%\uparrow$ & \textbf{92} & 61.4$\%\uparrow$ & \textbf{0.211~} & 101.1$\%\uparrow$ \\ 
\cline{2-13}
 & \multirow{3}{*}{CB} & \multirow{3}{*}{52} & CSN-t~ & 394,471 & \textbf{1} &  & 2 &  & 6 &  & 0.038~ &  \\
 &  &  & COFIC (controlled) & 394,471 & \textbf{1} & - & 3 & 50.0$\%\uparrow$ & 7 & 16.7$\%\uparrow$ & 0.043~ & 12.9$\%\uparrow$ \\
 &  &  & COFIC & 1,048,519 & \textbf{1} & - & \textbf{4} & 100.0$\%\uparrow$ & \textbf{7} & 16.7$\%\uparrow$ & \textbf{0.059~} & 56.3$\%\uparrow$ \\
\hline
\end{tabular}

%% file: threatstovalidity.tex
\section{Threats to validity}
\label{sec:threats}

\noindent\textbf{Rule Design} Though our experiments have evaluated the usefulness of each rule in the ruleset; the rule-based filter may still introduce a few false positives or false negatives due to its design and implementation. 
For example, the widely used query ``quick sort'' can be filtered out by the rule Short Sentences. 
Besides, some rules are tricky to be implemented exactly in line with our aim. 
For example, non-English letters in the comments are identified based on ASCII encoding. It may leave out several other languages also using English letters. But no English sentences will be falsely filtered out.
Overall, it requires further exploration on balancing the trade-off between precision and recall better.


\noindent\textbf{Bootstrap Query Corpus} The \corpus in this work is built based on the questions on StackOverflow. Only titles starting with ``how to'' are collected into the corpus, which limits the sentence pattern. The VAE model trained over this corpus may not have a good tolerance to other patterns.
Besides, StackOverflow titles are also not fully query-appropriate.
Although we filter the titles by rules, there are still semantically irrelevant texts left. 

\noindent\textbf{Generalization} Limited by the accessibility of models and evaluation benchmarks for code search tasks, we evaluate our solution only on Java datasets. In theory, our approach is capable of any comment-based code search dataset. Yet, the generalization of our filtering framework in different programming languages has not been experimentally verified. Besides, we only evaluate our filtering framework on two code search models, DeepCS and CARLCS, which is also a threat to the generalizability of our approach.

%% file: relatedwork.tex
\section{Related Work}
\label{sec:related}
\noindent\textbf{Code Search Dataset} Recent years have witnessed a growing interest in the semantic search for code snippets~\cite{kim2018facoy}. 
DL models are applied to establish links between natural language and programming language. To train these models~\cite{Gu2018DeepCS,Chen2018ANF,Cambronero2019WhenDL,Wan2019MultimodalAN,Yao2019CoaCorCA,Shuai2020ImprovingCS,Ling2020AdaptiveDC,Ye2020LeveragingCG,Wang2020TranS3AT,Haldar2020AMA,Li2020LearningCI,Qihao2020OCoRAO,Hu2020NeuralJA}, code snippets paired with comments are collected from Github~\cite{Barone2017APC,Gu2018DeepCS,Hu2018DeepCC,Wan2019MultimodalAN,Husain2019CodeSearchNetCE}. According to a manual investigation\cite{pascarella_classifying_2019}, there are 16 categories of comments in source code, most of which, e.g., TODO, License, and Exception, are not appropriate to serve as queries. However, to the best of our knowledge, the comments in code search datasets have never been fully cleaned. For example, Barone et al.~\cite{Barone2017APC} remove empty or non-alphanumeric lines from the docstrings. CodeSearchNet~\cite{Husain2019CodeSearchNetCE} filters each comment-code pair with its comment length. Ling et al.~\cite{Ling2020AdaptiveDC} use heuristic rules (e.g., the existence of verb and noun phrases) to filter comments. Cambronero et al.~\cite{Cambronero2019WhenDL} filter out queries that contain specific keywords. These simple and scattered efforts are not enough to filter out the various noises, especially the texts that are semantically unrelated to real queries. Liu et al.~\cite{Liu2020OpportunitiesAC} also mention that improving the data quality is still a research opportunity for deep-learning-based code search models, which well motivates our work.

There are two evaluation methods for neural code search research: train-test split and actual user query evaluation. A lot of works~\cite{Cambronero2019WhenDL, Shuai2020ImprovingCS,Wang2020TranS3AT,Ye2020LeveragingCG} split their datasets into train and test sets. The queries of their test set contain the same defects as the train set so that the results fail to reflect the model performance in an actual environment. 
Manually reviewed queries~\cite{Li2019NeuralCS,Husain2019CodeSearchNetCE,Yan2020AreTC} can overcome this problem but they are usually on a small scale and cannot serve as the training dataset. 


\noindent\textbf{Unsupervised Anomaly Detection} Comments cleaning is an application of the unsupervised anomaly detection algorithm as labeled comments are non-trivial to obtain. Unsupervised anomaly detection algorithms identify the outliers solely based on the intrinsic properties of the data instances. Various techniques can be applied, such as  Principal Component Analysis~\cite{Wold1987PrincipalCA}, Generative Adversarial Network~\cite{Lawson2017FindingAW}, Spatio Temporal Networks~\cite{Chianucci2016UnsupervisedCD} and LSTM~\cite{Singh2017AnomalyDF}. Among them, Auto-Encoder (AE) is the fundamental architecture for unsupervised anomaly detection~\cite{Sydney2019DeepLF}. It has been applied in many tasks. For example, Zhang et al.~\cite{Zhang2017DetectingRO} detect the rumors in social media using multi-layer AE. Castellini et al.~\cite{Castellini2017FakeTF} apply AE to detect false followers on Twitter. Luo and Nagarajan~\cite{Luo2018DistributedAD} use AE to identify the error events of interest such as equipment faults and undiscovered phenomena in wireless sensor networks. 

The encoder of AE maps an input to a point in the latent space, while VAE maps an input to a region. In this way, VAE can extract more abstract semantic features. It has been applied to unsupervised anomaly detection with promising evaluation scores~\cite{An2015VariationalAB,Suh2016EchostateCV, Xu2018UnsupervisedAD, Lin2020AnomalyDF}.